

\input amstex.tex

\documentstyle{amsppt}
\magnification\magstep1
\topmatter\title Current algebras in $3+1$ dimensions\endtitle
\author Jouko Mickelsson\endauthor
\affil  Department of Mathematics, University of
Jyv\"askyl\"a, SF-40100, Jyv\"askyl\"a, Finland \endaffil
\endtopmatter
\document

\redefine\e{\epsilon}

\redefine\o{\omega}

\redefine\l{\lambda}

\define\CM{\Bbb C}

\define\gm{\bold g}

\define\<#1,#2>{\langle #1,#2\rangle}
\define\TR{\text{tr}}
\define\dep(#1,#2){\text{det}_{#1}#2}

\NoBlackBoxes
ABSTRACT Aspects of a generalized representation theory of current algebras
in $3+1$ dimensions are discussed in terms of the Fock bundle method, the
sesquilinear form approach (of Langmann and Ruijsenaars), and Hilbert space
cocycles.

\define\GL{\widehat{GL}}
\vskip 0.3in

Let $G$ be a Lie group of symmetries of a classical relativistic field theory.
If $\{T_1,\dots,T_N\}$ is a basis for the Lie algebra $\gm$ of $G$ then to each
$T_a$ there corresponds, by Noether's theorem, a conserved current
$j^a_{\mu}(x). $ (Actually, the current should be thought of as a 3-form.)

When the classical field theory is quantized (in the Hamiltonian approach) on a
space-like surface $S$ then to each current $j^a$ is supposed to correspond an
operator valued distribution on the surface $S.$ Formally,
$$j^a(f)=\int_S f(x) j^a(x),\tag1$$
where $f$ is a smooth test function. Naively, one would expect that the charges
$j^a(f)$ satisfy the commutation relations
$$[j^a(f),j^b(g)]=\l^{ab}_c j^c(fg),\tag2$$
where the $\l^{ab}_c$'s are the structure constants,
$[T_a,T_b]=\l^{ab}_c T_c.$

However, in order to make sense of the quantum operators $j^a(f)$ one normally
has to do certain regularizations. For example, in a field theory involving
fermions in $1+1$ space-time dimensions, transforming according to a
finite-dimensional representation $\rho$ of $G,$ the group generated by the
charge densities is the group $Map(S,G)$ of smooth maps from $S$ to $G,$
acting on the fermion field $\psi$ by pointwise multiplications $\psi'(x)=
\rho(f(x))\psi(x).$ In the second quantized theory the charge densities are
given naively by the operators
$$T(f)=\sum f_{nm} a^*_n a_m,\tag3$$
where the $a^*_n$'s are the fermion creation operators and the $a_m$'s are
annihilation operators; here $f_{nm}$ denote the matrix elements of the
multiplication operator in a given orthonormal basis in the one-particle space.

The problem with the the above construction is that in general a state created
from the vacuum would have infinite norm. For making the operators finite it is
sufficient to make a normal ordering: The products $a^*_n a_m$ are replaced by
$:a^*_n a_m:$ where the dots mean that the operators lowering the energy are
placed to the right. The normal ordering changes the commutation relations
(2). Let $\pi_+$ be the projection onto the positive energy subspace in the
one-particle Hilbert space $H$ and $\epsilon=2\pi_+-1.$ Denoting by $X,Y$
components of the charges in the one-particle representation, the new
commutators are
$$[X,Y]'=[X,Y]+ \frac14\TR \epsilon[[\epsilon,X][\epsilon,Y].\tag4$$

The second term on the right in (4) is an antisymmetric bilinear form $c$ on a
Lie algebra \define\gl{\bold{gl}} $\gl_1$ which contains the algebra of
integrated charge densities $j^a(f).$ The algebra $\gl_1$ consists of bounded
operators $X$ in $H$ such
that $[\epsilon,X]$ is a Hilbert-Schmidt operator. The form $c$ (Lundberg's
cocycle, [Lu]) satisfies the
\it cocycle condition \rm
$$c(X,[Y,Z])+c(Y,[Z,X])+c(Z,[X,Y])=0 \tag5$$
which guarantees that the Jacobi identity for the modified commutators $[\cdot,
\cdot]'$ is satisfied.

In quantum field theory in space-time dimensions higher than two the normal
ordering of the charge densities is not sufficient to make them finite; more
radical regularizations are needed. In fact, there is no meaningful way to
make the $j^a(f)$'s into operators in the (fermionic or bosonic) Fock space.
The technical reason for this is that the currents are no more contained in
$\gl_1$ but in a bigger Lie algebra $\gl_p,$ where $p$ must be greater or equal
to $\frac12(d+1),$ where $d$ is the space dimension and $\gl_p$ is
characterized
by the property that $[\epsilon,X]$ belongs to the Schatten ideal $L_{2p}$
consisting of operators $T$ such that $(T^*T)^p$ has a finite trace. Note
that the cocycle $c$ is defined only on $\gl_1$ and not on any of the
larger algebras $\gl_p,$ $p>1.$

There is a modified cocycle $c_p$ which converges for $\gl_p.$ The coefficient
ring for these cocycles is not $\CM$ but a certain space of complex valued
functions on an infinite-dimensional manifold (which in this talk will be
a Grassmannian manifold). For example, when $d=3$ the cocycle takes the form,
[MR],
$$c_2(X,Y;F)=\frac18\TR(\e-F)[[\e,X],[\e,Y]],\tag6$$
where $F$ is a hermitean operator, with $F^2=1$ and $\e F+F\e-2\e\in L_2,$
parametrizing points on a Grassmannian $Gr_2.$ For $p>2$ the cocycles $c_p$
have been computed in [FT].

The new Lie algebra $\widehat{\bold{gl}}_2$ defined by the cocycle (6) does
not have interesting Hilbert space representations, [P]. Thus we have to look
for a generalization of representation theory.

Because of the "background field" $F$ the current algebra acts in a bundle of
Fock spaces parametrized by $Gr_2,$ [M1]. There is an alternative description
of the generalized representation of the current algebra proposed by
E. Langmann [L]. The starting point is the observation of S.N.M. Ruijsenaars
that although the charges are not well-defined operators in the Fock space,
they
still make sense as sesquilinear forms, [R1]. The problem with sesquilinear
forms
is that one cannot multiply them like matrices; if one tries to do that one
gets
in general a divergent expression. Langmann showed that there is a regularized
multiplication of the forms which leads to the cocycle (6).  In fact, one
can work backwards from the algebra to construct the regularized product,
[M2].

A computation of the vacuum expectation value (with respect to the free vacuum
parametrized by $F=\e$) of a product of operators $X_1 X_2\dots X_n$
corresponding
to charges (1) is now completely algebraic. Roughly speaking, any $X_i$ can be
split into an energy increasing operator $X_+$ and an energy lowering operator
$X_-.$ When acting on the vacuum on the right $X_-$ gives zero whereas $X_+$
hitting the vacuum on the left in $<0|X_1\dots X_n|0>$ gives zero. Using
repeatedly the commutation relations (6) one reduces the operator product to
a product of the cocycles $c_2$ (which are functions of $F$) sandwiched
between the vacua. As a final step one evaluates all cocyles at $F=\e,$ see
[M2] for details.

In addition to the description of the current algebra as Fock bundle maps or
as sesquilinear forms in a fixed Fock space there is a third alternative way
for constructing a "representation". Let $\Cal F_A$ be the family of
(fermionic) Fock spaces parametrized by the space $\Cal A$ of external vector
potentials $A.$ Since $A$ is flat the Fock bundle $\Cal F$ is trivial; choose
a trivialization by choosing for every $A\in \Cal A$ a unitary operator
$h_A:H_0\to H_A,$ where $H_A$ is the one-particle Hilbert space corresponding
to the potential $A.$ The operator $h_A$ should satisfy the condition that
$\epsilon_A h_A -h_A\epsilon_0$ is Hilbert-Schmidt; here $\e_A=2 \pi_{+,A}-1,$
$\pi_{+,A}$ is the projection onto the positive energy subspace in the
background $A.$ If $S\in GL_1$ belongs to the unitary subgroup $U_1$ then also
$h_A S$
satisfies the same condition. On the other hand, if both $h_A$ and $h$ satisfy
the condition above, then $h=h_A S$ for some $S\in U_1.$
It follows that the set of allowed $h_A$'s form a principal bundle $P$ over
$\Cal A$ with structure group $U_1.$ Since the base space $\Cal A$ is flat, the
bundle has a smooth section $h_A.$

Denote $D^h_A=h_A^{-1}D_A h_A.$ By the basic property of $h_A$ the self-adjoint
operator $D^h_A$ is quantizable in the free fermionic Fock space $\Cal F_0,$
and for each $\l$ not in the spectrum of $D_A$ there is  a unique vacuum ray
$|A,\l>$ characterized by
$$a^*(v)|A,\l>=0=a(u)|A,\l>$$
where $v$ (resp. $u$) is an eigenvector of $D^h_A$ belonging to an eigenvalue
smaller (resp. bigger) than $\l.$ Thus we have a trivial bundle of Fock spaces
parametrized by the potentials. Note however that there is no continuous global
section $\psi$ such that $\psi(A)$ is a vacuum for $D^h_A.$ This is because
the eigenvalues of the Dirac operator, as a function of $A$, can cross any
given vacuum level $\l.$

The action of a gauge transformation $g$ is given as follows. Besides the
usual gauge action $A\mapsto g\cdot A=gAg^{-1}+dgg^{-1}$ in the base space
$\Cal A$ we have an action in the fibers
of the Fock bundle, which is obtained by quantizing the one-particle operator
$T_A(g)=h_{g\cdot A}^{-1}T(g)h_A.$ The quantization of the operator $T_A(g)\in
GL_1$ is naturally defined only up to a phase (because of the nontrivial
central extension $\GL_1$). Again since $\Cal A$ is flat, the phase for any
given
$g$ can be chosen as a continuous function of $A.$ However, as a function of
$g$
the choice cannot be continuous (topological obstruction from the nontriviality
of  $\GL_1\to GL_1$). If we have made a choice of the phase
for three elements $g_1,g_2,$ and $g_1g_2$, then we can write
$$\hat T_A(g_1g_2)=\o(g_1,g_2;A)\hat T_A(g_1)\hat T_{g_1^{-1}\cdot A}(g_2),$$
where the phase $\o$ is also a function of the base point $A.$

Thus, although we still do not have a normal Hilbert space representation
of the current algebra, we have a \it cocycle valued in the representation
of the smaller algebra \rm $\GL_1.$ Details of the construction will appear
elsewhere.

In $1+1$ dimensions there is a famous "Bose-Fermi correspondence" based on
representation theory of Kac-Moody algebras. There
is a  modification of the
Bose-Fermi correspondence which works also in $3+1$ space-time dimensions.
The first indication in this direction was the 'kink' construction of fermions
by Finkelstein and Rubinstein, [FR]. For example, take
$M=S^3$ and $G=SU(N).$ A kink is then a noncontractible map $f:S^3\to G.$  The
homotopy classes of kinks are classified by an element of $\pi_3 G,$ which is
equal to $\Bbb Z$ when $N\geq 2.$ When $N=2$ the fourth homotopy group
$\pi_4 G$ is $\Bbb Z_2.$ In [FR] it was shown that in a configuration
containing two nonoverlapping kinks a continuous rotation of one of the kinks
is homotopic to a continuous interchange of the kink positions.
This is a kind of spin-statistic theorem (later generalized
by Sorkin, [So]). In particular, the case $G=SU(2)$ allows two types of field
configurations. A contractible loop in $\Cal G=Map(S^3,G)$ is interpreted
as a boson whereas a noncontractible loop (generator of $\pi_4 SU(2)$) is
a fermion.

The possibility of realization of fermions in terms of seemingly bosonic
fields really occurs in the WZW model, as shown by Witten in [W]. The
quantum mechanical WZW action functional obtains a factor $-1$ each time
as the field $f:S^3\to SU(n)$ is adiabatically rotated by the angle $2\pi.$
On the other hand, it was proved in [M3] that if $\Cal G$ acts projectively in
the quantum vector space (the nontrivial projective phases are related to
chiral anomaly) then $D(f)D(g)=(-1)^{n(f)n(g)} D(g)D(f),$ where $n(f)$ is the
winding number of $f.$ I want to stress
that this relation does not depend on the existence of a unitary (projective)
representation $D$ (that may not exist) in the ordinary sense. The group
has nonunitary linear representations which have some similarities with
the unitary highest weight representations in the one-dimensional case, [MR].
The formula is purely a result of the structural relations in an extension
$\hat{\Cal G}.$

A more recent and important result on the Bose-Fermi relation in $3+1$
dimensions is due to Ruijsenaars, [R2]. As we have seen, although the operators
$D(g)$ representing gauge transformations do not exist as linear operators in
the fermionic Fock space, they still exist as sesquilinear forms in a suitable
dense domain. Furthermore, one can take limits of kinks $f_n: S^3\to G$
such that lim$\,D(f_n)$ exists as a bilinear in the Fock space, when
lim$\,f_n$ is a 'pointlike kink', the winding being concentrated to a single
point $x\in S^3.$ The limiting bilinear is the fermion field.

\vskip 0.1in
\bf REFERENCES \rm

\vskip 0.1in

\bf [FR] \rm Finkelstein, D. and J. Rubinstein. J. Math. Phys. \bf 9, \rm 1762
(1968). \bf [FT] \rm Fujii, K., and M. Tanaka. Commun. Math. Phys. \bf 129,
\rm 267 (1990).
\bf [L] \rm Langmann, E. Proc. of the colloquim "Topological and Geometrical
Methods in Field
Theory", Turku, Finland, May 1991 (eds. by J. Mickelsson and O. Pekonen,
World Scientific, Singapore, 1992).
\bf [Lu] \rm Lundberg, L.-E. Commun. Math. Phys.
\bf 50, \rm 103 (1976).
\bf [M1] \rm Mickelsson, J. Commun. Math. Phys. \bf 127, \rm 285 (1990);
\bf [M2]  \rm -------- M.I.T. preprint CTP\#2107, June 1992;
\bf [M3]  \rm --------: \it Current Algebras and Groups. \rm Plenum Press,
New York and London (1989).
\bf [MR] \rm Mickelsson, J. and S. Rajeev. Commun.
Math. Phys. \bf 116, \rm 365 (1988).
\bf [P] \rm Pickrell, D. Commun. Math. Phys. \bf 123, \rm 617 (1989).
\bf [R1] \rm Ruijsenaars, S.N.M. J. Math. Phys. \bf 18, \rm 517 (1977);
\bf [R2] \rm ------- Commun. Math. Phys. \bf 124, \rm 553 (1989).
\bf [So] \rm Sorkin, R.\nopagebreak Commun. Math. Phys. \bf 115, \rm 421
(1988).
\bf [W] \rm Witten, E. Nucl. Phys. \bf B223, \rm 433 (1983).

\enddocument